\documentclass[aps,prl,twocolumn,groupedaddress,showpacs]{revtex4}
\usepackage{graphicx}
\newcommand{\gma}{$\rm{Ga_{1-x}Mn_{x}As}$}
\newcommand{\rxx}{$\rho_{\rm{xx}}$}
\newcommand{\rxy}{$\rho_{\rm{xy}}$}
\newcommand{\sxx}{$\sigma_{\rm{xx}}$}
\newcommand{\sxy}{$\sigma_{\rm{xy}}$}

\newcommand{\tc}{$T_{\rm{C}}$}

\begin{document}

\title{Localization and the Anomalous Hall Effect in a ``Dirty" Metallic Ferromagnet}

\author{P. Mitra}
\author{N. Kumar}
\author{N. Samarth}
\email{nsamarth@phys.psu.edu}
\affiliation{Department of Physics, The Pennsylvania State
University, University Park, PA 16802}

\date{\today}

\begin{abstract}

We report magnetoresistance measurements over an extensive
temperature range (0.1 K $\leq T \leq$ 100 K) in a disordered
ferromagnetic semiconductor (\gma). The study focuses on a series of
metallic \gma~ epilayers that lie in the vicinity of the
metal-insulator transition ($k_F l_e\sim 1$). At low temperatures ($T
< 4$ K), we first confirm the results of earlier studies that the
longitudinal conductivity shows a $T^{1/3}$ dependence, consistent
with quantum corrections from carrier localization in a ``dirty''
metal. In addition, we find that the anomalous Hall conductivity
exhibits universal behavior in this temperature range, with no
pronounced quantum corrections. We argue that observed scaling relationship between the
low temperature longitudinal and transverse resistivity, taken in conjunction with the absence
of quantum corrections to the anomalous Hall conductivity,  is consistent with the side-jump mechanism for the anomalous Hall
effect. In contrast, at high temperatures ($T \gtrsim 4$ K), neither
the longitudinal nor the anomalous Hall conductivity exhibit
universal behavior, indicating the dominance of inelastic scattering
contributions down to liquid helium temperatures.
\end{abstract}


\maketitle

\section{Introduction}
The anomalous Hall effect (AHE) refers to an excess transverse Hall
voltage arising from the influence of spin-orbit coupling on the
motion of charge carriers in a spin polarized metal. The phenomenon
first attracted attention several decades ago
\cite{KLPhysRev.95.1154,SmitPhysica.21.877,kondo:ah,BergerPhysRevB.2.4559}
and still continues as an active area for theoretical
\cite{nagaosa:ah,muttalib:214415,onoda:126602,DugaevPhysRevB.64.104411,JungwirthPhysRevLett.88.207208,nagaosa-2009}
and experimental
\cite{chun:026601,cumings:196404,pu:117208,Tian:2009yf,Chiba:2010vl,nagaosa-2009}
inquiry.  Calculations have identified two classes of underlying
mechanisms for the AHE: ``extrinsic'' contributions originating from the
spin-dependent scattering of carriers by impurities
\cite{SmitPhysica.21.877,BergerPhysRevB.2.4559} and ``intrinsic''
ones that involve the interaction of carrier spins with the inherent
crystal band structure \cite{KLPhysRev.95.1154,zhang:PRL}.  A number
of experiments have aimed to identify the mechanisms responsible for
the AHE in different materials (see ref. \cite{nagaosa-2009} for a
complete bibliography). This is commonly addressed by examining a
scaling relationship of the form $\rho_{\rm{xy}}\sim
\rho_{\rm{xx}}^\alpha$, where \rxx~and \rxy~are the longitudinal
and transverse resistivity, respectively. Theory predicts $\alpha =
2$ for the intrinsic mechanism \cite{KLPhysRev.95.1154}. In the
extrinsic case, the exponent depends on the nature of the spin
dependent scattering mechanism: $\alpha=1$ for skew scattering
\cite{SmitPhysica.21.877} and $\alpha=2$ for the side jump mechanism
\cite{BergerPhysRevB.2.4559}.

Three regimes emerge from these experimental studies \cite{nagaosa-2009}: \begin{itemize}
\item at low disorder, the scaling exponent $\alpha =1$, suggesting that the AHE is dominated by skew scattering;
\item at moderate disorder, $\alpha =2$, resulting in a Hall conductivity \sxy~ which is independent of disorder (since $\sigma_{\rm{xy}} = \frac{\rho_{\rm{xy}}}{\rho_{\rm{xx}}^2}$). This scaling is consistent with both the intrinsic mechanism and the extrinsic side-jump mechanism, but it is difficult to convincingly distinguish between these using dc transport;
\item at large disorder, numerical calculations report $\alpha \approx 1.6$. While there is no simple physical explanation for this behavior, this scaling does appear to emerge from numerical calculations of the anomalous conductivity in the hopping regime \cite{onoda:126602}. 
\end{itemize}
A cautionary note arises in the interpretation of such experimental
data: the analysis of the scaling relationship should strictly be
carried out at temperatures where impurity scattering dominates and
other inelastic processes (such as scattering from spin waves) have
been frozen out.

In this paper, we describe measurements of the AHE in a canonical
ferromagnetic semiconductor (\gma) over a large temperature range
(100 mK $\leq T \leq $150 K). Our study is restricted to samples
whose resistivity lies in a regime 2 $\rm{m}\Omega . \rm{cm} \lesssim
\rho_{\rm{xx}} \lesssim$ 10 $\rm{m}\Omega . \rm{cm}$ where other
studies \cite{chun:026601,pu:117208} have observed an AHE scaling
exponent $\alpha = 2$ at $T \geq 4.2$ K. These observations have
been interpreted as an insensitivity to disorder and thus as a
signature of the intrinsic mechanism \cite{chun:026601}, in
agreement with theoretical predictions made in the disorder-free
limit \cite{JungwirthPhysRevLett.88.207208}. However, we argue that
unlike metallic ferromagnets where the Drude approximation works well at
helium temperatures, the complex non-monotonic temperature dependence
of the resistivity in \gma~provides strong motivation for 
systematic measurements at lower (dilution fridge) temperatures
where impurity scattering is dominant. Our principal aim is to test
the robustness of the scaling relationship by examining it at such
lower temperatures. Our study also allows us to address another
interesting question that has thus far been ignored in \gma: what
(if any) are the quantum corrections to the anomalous Hall
conductivity? This question is particularly germane within the
context of \gma~where the confluence of ferromagnetism, disorder
and interactions invariably occurs in close vicinity of the
metal-insulator transition (MIT) \cite{Richardella:2010hb}.  Earlier
studies \cite{muttalib:214415,mitra:046804} on metallic ferromagnets in the weak disorder limit ($k_F
l_e \gg1$, where $k_F$ and $l_e$ are the Fermi wave vector and the elastic mean free path, respectively) show that the quantum
correction to the anomalous Hall conductivity is distinctly
different for the skew scattering and side jump mechanisms, thus
providing an alternative route to test the origin of the AHE. We are however unaware of any theoretical calculations addressing the corresponding quantum correction for the intrinsic mechanism.

\section{Experimental Details}
Our measurements center on a set of five \gma~samples grown by
low temperature molecular beam epitaxy,
with Mn composition in the range $0.028 \leq x \leq 0.078$.  The samples are all 120
nm thick, with the exception of the sample with $x=0.078$ which is 50 nm
thick. A detailed description of the growth and characterization of
these samples has been provided earlier
\cite{potashnik:1495,ku:2302}. In order to generate a larger phase
space of conductivity, we measure both as-grown and annealed pieces
of these samples. The annealing is known to simultaneously increase the hole density and decrease the disorder by reducing the density of hole-compensating Mn interstitial defects \cite{potashnik:1495,ku:2302}. Transport measurements are carried out on
lithographically defined Hall bars (400 $\mu$m $\times$ 100 $\mu$m) in a
Quantum Design dilution refrigerator with the sample temperature ranging from 100
mK to 4 K. (Although the nominal base temperature of the dilution fridge is 50 mK,
we find that samples do not cool down efficiently below 100 mK and thus exclude
the corresponding data from our analysis.)  We measure longitudinal and transverse resistances
simultaneously using a standard double lock-in technique with an
a.c. excitation of 100 nA (optimal for measuring the small Hall
signal while minimizing sample heating). To avoid contributions to
the small Hall voltage from the significantly larger longitudinal
voltage, we carry out measurements at positive and negative applied
magnetic field $B=0.5$ T (sufficient to fully polarize magnetic
domains) and use the standard anti-symmetrization procedure. We note
that -- due to the large hole density ($p \geq 10^{20}
\rm{cm}^{-3}$) -- the contribution of the ordinary Hall effect is
negligible compared to the AHE. Since our analysis
relies on knowledge of the hole carrier density, we also note that the values
used here are determined using Raman measurements \cite{PhysRevB.66.033202}, reducing the uncertainties that are typically involved in deducing the carrier density in the presence of the AHE. 

\section{Quantum Corrections to the Longitudinal Conductivity}
\begin{figure}
\includegraphics[width=8cm]{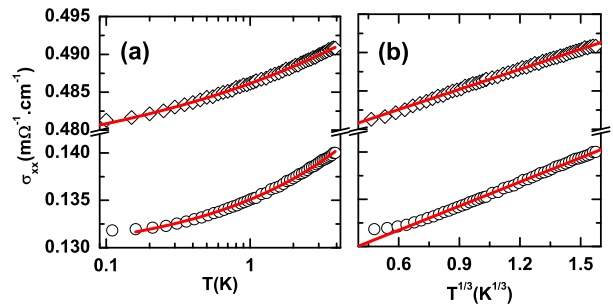}
\caption{\label{Figure1}(a) Plot of conductivity vs. temperature on a
log scale for as grown $x=.03$ (diamonds) and annealed $x=.08$ (circles) \gma. Solid lines represents fits to
$\sigma_{xx}(T)=\sigma_{xx} ^0+AT^n$. (b)Plot of conductivity of the
same samples vs. $T^{1/3}$. Solid lines are linear fits.}
\end{figure}

We first address the temperature dependence of the zero
field conductivity (\sxx=1/\rxx) shown in Fig.
\ref{Figure1}(a) for
two extremes in our sample set: the as grown
sample with $x = 0.028$ has the lowest value of \sxx, while the
annealed sample with $x = 0.078$ has the highest \sxx. An unbiased fit
of the data to the functional form:
 $\sigma_{\rm{xx}}(T)=\sigma_o+AT^n$, with the exponent $n$ as a free
fitting parameter, yields $0.2 \lesssim n \lesssim 0.4$ for
different samples.  Based on the known functional forms for the
temperature dependence of \sxx~in three dimensional (3D) disordered
systems, combined with earlier analyses of \gma~
\cite{honolka:245310}, $\rm{In_{1-x}Mn_{x}As}$~
\cite{PhysRevLett.68.2664} and (non-magnetic) n-GaAs
\cite{capoen:2545}, we conclude that our data is most appropriately
described by $n=1/3$.  Indeed, a plot of \sxx~vs. $T^{1/3}$ shows a
linear dependence over almost two decades of variation in $T$ (Fig.
\ref{Figure1}(b)). As we discuss below, the  $T^{1/3}$ dependence of
conductivity in 3D disordered systems has been predicted within the
context of the scaling theory of localization in the ``dirty metal"
regime \cite{Althsuler:1983zp}. We note that other studies of the
temperature-dependent conductivity in \gma~have taken a different
interpretation, assigning a \sxx $\sim T^{1/2}$ dependence which
arises from electron-electron interactions \cite{Neumaier:2009pt}.
However, forcing a $T^{1/2}$ fit to our data yields a systematic
variation in the residual plot, leading us to discount this
dependence as providing a physically meaningful description of the
data. We also note that the $\sim T^{1/2}$ dependence is derived
from perturbative calculations that apply only to weakly disordered
systems ($k_F l_e >> 1$), while even the best \gma~samples
fabricated to date are highly disordered and always close to the
MIT. For the samples studied here, we estimate Drude values of $l_e
\sim 0.5-1.3$ nm and $k_F l_e \sim1-2$ using the extrapolated zero
temperature values of \sxx.

As first pointed out by Altshuler and Aronov
\cite{Althsuler:1983zp}, in the vicinity of the MIT, the
disorder-dependent correlation length is larger than relevant
temperature-dependent length scales, resulting in a temperature
variation of the scale dependence of the diffusion constant $D$
\cite{imry:1817}. This contrasts with the weak disorder regime where
$D$ is independent of temperature. One possibility is that the dominant temperature
dependent length scale is the interaction length defined as $L_T=\sqrt{\hbar
D/k_BT}$: physically, this characterizes the length over which coherence is maintained during the elastic scattering of quasiparticles that lie within $k_B T$ of the Fermi energy. In this case \cite{PhysRevLett.61.369}, the prefactor of $T^{1/3}$
is proportional to the density of states $N$; specifically,
$A=(e^2/\hbar)(G_c^2Nk_B)^{1/3}$, where $G_c$ is the critical
conductance and $k_B$ is the Boltzmann constant. Another possibility is that the dominant
length scale is the inelastic length $l_{in}=\sqrt{D\tau_\phi}$ which characterizes the diffusive motion of electrons between inelastic collisions \cite{ImryOvadyahu1982}. Here, the 
dephasing rate is $\tau_{\phi}^{-1}=aT$ and the prefactor of $T^{1/3}$
is $A=(e^2/\hbar)(G_cNak_B)^{1/3}$. A relevant dephasing
mechanism in ``dirty metals'' is that due to the electron-electron
interaction \cite{isawa:1984}; in this case, $a\sim(k_F l_e)^{-2}$.
Hence, the prefactor $A$ should depend on both disorder and carrier
density. We note that in \gma~other possible dephasing mechanisms
may also exist, including scattering from spin waves or from two
level systems formed at Mn interstitial sites \cite{Zhu:2007fz}.
However, we do not anticipate an explicit dependence on disorder in
the latter cases.

\begin{figure}
\includegraphics[width=8cm]{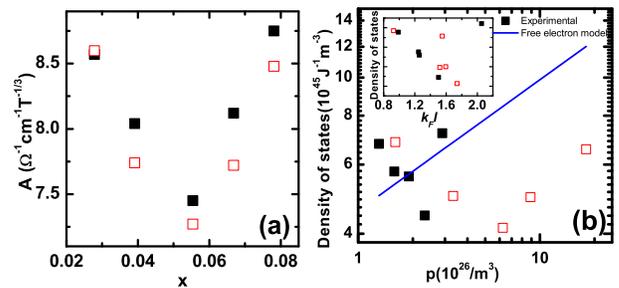}
\caption{\label{Figure2}(a) Prefactor A vs. Mn concentration obtained
from numerical fits of the conductivity data to
$\sigma_{xx}(T)=\sigma_{xx} ^0+AT^{1/3}$. (b) Experimental estimate of the
density of states $(\hbar A/e^2)^3/k_{\rm{B}}$ vs. hole density. The solid line shows the expected value for a free electron gas.
Inset shows the density of states versus
$k_F l_e$. The filled and open squares represent as-grown and annealed
samples, respectively.}
\end{figure}

Figure \ref{Figure2}(a) plots the prefactor $A$ as a function of the
Mn concentration ($x$) for both as-grown and annealed samples,
revealing a non-monotonic dependence wherein $A$ reaches a minimum
at $x\sim 0.056$. Since Mn acts both as an acceptor that generates
holes and also adds to the disorder in the form of interstitials, we
speculate that -- for given growth conditions -- there exists an
optimal Mn doping level that maximizes the carrier density and
minimizes disorder. Further -- for a given Mn concentration -- $A$
is found to decrease upon annealing which is known to increase the
hole carrier density. This is inconsistent with the interaction
picture that predicts $A$ to increase if $p$ increases. Thus, the
data in Fig.\ref{Figure2}(a) indirectly suggest that $A$ varies with
disorder and that the $T^{1/3}$ dependence is possibly due to
localization rather than interactions.

Recent experiments on mesoscopic \gma~samples
\cite{Wagner:2006kk,vila:027204} estimate inelastic path lengths
$l_{in}\sim100$ nm  for $x=0.02$ at 10 mK and for $x=0.06$ at 100
mK, respectively, and a dephasing time $\tau_{\phi}$ that varies
inversely with temperature at higher temperatures ($T \lesssim 1$ K). This places our samples of thickness $t=120$ nm in the 3D
regime (wherein the film thickness exceeds the inelastic length
$l_{in}$) in the temperature range of interest (100 mK$ \leq T
\leq 4$ K). Further, if we compare these estimates for the
inelastic path lengths with our Drude estimates of the elastic path
length, we find that the condition $l_{in}> l_e$ is indeed satisfied
in our samples. Figure \ref{Figure2}(b) plots the quantity $N=(\hbar
A/e^2)^3/G_c^2k_B$, as a function of $p$ and $k_F l_e$(inset), where
we use $G_c=3\pi^3/2$ \cite{PhysRevLett.61.369}. Note that in the
interaction picture \cite{PhysRevLett.61.369}, $N$ is indeed the
density of states while in the localization
case \cite{ImryOvadyahu1982}, $N$ depends on both density of states
and disorder. As shown in Fig.\ref{Figure2}, $N$ does not follow any
systematic dependence on $p$ and decreases with increasing $k_F
l_e$(inset), again supporting the localization picture (the $t=50$
nm samples do not follow the trend).

\section{Absence of Quantum Corrections to the Anomalous Hall Conductivity}
\begin{figure}
\includegraphics[width=8cm]{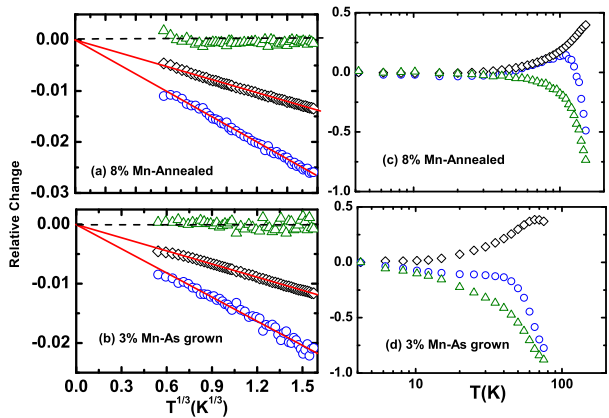}
\caption{\label{Figure3} Relative changes in the longitudinal
resistivity (diamonds), the anomalous Hall resistivity (circles) and
the anomalous Hall conductivity (triangles) for two \gma~ samples.
In (a) and (b), these are plotted as a function of $T^{1/3}$ for 0.1 K $\leq T \leq $ 4 K. Solid
lines are linear fits to the data. In (c) and (d), these quantities are plotted vs. $T$ for $T \geq 4$ K.}
\end{figure}

We now address the quantum corrections to  \sxy~by using the temperature variation of
\rxx~and \rxy.  We find that for
all our samples, both \rxx~and \rxy~increase
monotonically with decreasing temperature in the range 100 mK$
\leq T \leq 4$ K and we observe that the temperature dependence can be fit
to the functional form $\rho_{ij}(T)=\rho_{ij}^0 + A_j T^{1/3}$. Thus
at $T=0$ K, the resistivity extrapolates to a finite  $\rho_{ij}^0$ and  is
found to be positive for both longitudinal and Hall components for
each sample. To
estimate the magnitude of the quantum corrections, we examine the
relative changes of the resistivity with respect to the zero temperature
value ($\delta_T\rho_{ij}=(\rho_{ij}(T)-\rho_{ij}(0))/\rho_{ij}(0)$.
We also examine the relative change in the anomalous Hall
conductivity calculated using
$\sigma_{\rm{xy}}=\rho_{\rm{xy}}/(\rho_{\rm{xx}}^2+\rho_{\rm{xy}}^{2})$. Figures
\ref{Figure3}(a) and (b) plot $\delta_T\rho_{\rm{xx}}$,
$\delta_T\rho_{\rm{xy}}$ and $\delta_T\sigma_{\rm{xy}}$ as a function of
$T^{1/3}$. Linear fits to the data indicate that the slope
$\delta_T\rho_{\rm{xy}}$ is double that of $\delta_T\rho_{\rm{xx}}$. Also,
$\delta_T\sigma_{\rm{xy}}$ does not show any pronounced temperature
dependence and is scattered around zero over the temperature range.
Using the approximation $\sigma_{\rm{xy}}\approx\rho_{\rm{xy}}/\rho_{\rm{xx}}^2$,
it is easy to show that the following identity summarizes the
universal scaling behavior observed in all our samples:
\begin{equation}
\delta_T\rho_{\rm{xy}}=2\delta_T\rho_{\rm{xx}}
\Longrightarrow\delta_T\sigma_{\rm{xy}}=0
\end{equation}
Thus, we report a key experimental finding: there are no quantum
corrections to $\sigma^{AH}_{xy}$~ in metallic \gma~ over the
temperature range of 100 mK $\leq T \leq $4 K where \sxx~
exhibits a finite localization correction in the form of $T^{1/3}$.
We are not aware of any theoretical calculations that support our
findings in this high disorder regime. However, we note that
calculations of quantum corrections to the anomalous Hall
conductivity in {\it weakly disordered} ferromagnets
\cite{DugaevPhysRevB.64.104411,muttalib:214415} do indeed show that
-- in contrast to the skew scattering mechanism -- the side jump
mechanism results in a negligible temperature dependence of \sxy.
Speculating that the essential nature of this low disorder result is
unchanged at high disorder, our observations could be consistent
with the side jump mechanism.

\section{Behavior of Longitudinal and Anomalous Hall Conductivity at High Temperatures}

To further corroborate the uniqueness of the low temperature results
described above, we carried out measurements on our samples at
higher temperatures ($T \geq 4.2$ K), up to the Curie temperature
\tc~of the samples. Not surprisingly, we find that the universal
scaling of \rxx~and \rxy~(which leads to a constant \sxy~at low
temperatures) breaks down at higher temperatures. As is well known,
the temperature dependence of \rxx~shows a peak near \tc~ and
decreases to a minimum around $T\sim 8-15$~K, showing a much weaker
monotonic increase with decreasing temperature. The temperature
dependence of \rxx~for $T \geq 4$ K has recently been interpreted
within a scaling analysis picture that takes into account the
competition between localization and the onset of magnetic ordering
\cite{moca:137203}. However, the temperature dependence of \rxy~and
\sxy~has not been studied previously in great detail over a wide
range of samples. We find  a non-universal behavior in the relative
scaling of \rxx~and \rxy~at high temperature. Figures
\ref{Figure3}(c) and (d) plot the temperature-dependence of the
relative changes in \rxx, \rxy~and \sxy~with respect to their values at $T=4.2$~K. For low Mn content samples, \rxy~decreases monotonically with increasing temperature in the regime
4.2 K$\leq T \leq$ \tc, as shown in Fig.\ref{Figure3}(d). For higher
Mn content samples, the temperature variation of \rxy~is
non-monotonic over this temperature range. As shown in
Fig.\ref{Figure3}(c), the relative changes in \rxy~and \rxx~increase with the same rate up to $T\sim 90$~K; at even higher
temperatures, \rxy~decreases rapidly as \tc~is approached. In both
cases, magnetometry measurements (data not shown) show that the saturated magnetization $M_s$ decreases
monotonically as $T$ increases. The observations are consistent with
the view that in low  \tc~ferromagnets such as \gma, magnetic
fluctuations play an important role at high temperatures ($T \gtrsim
8$ K) and dominate the temperature dependence of the resistivity
\cite{moca:137203}. In such regimes, scattering from magnetic
impurities may contribute to the AHE with \rxy~proportional to
magnetic fluctuations of the form $\langle (M - \langle M \rangle)^3 \rangle$. We speculate that
this ``Kondo-type"  AHE may be more relevant at high temperatures.

In this high temperature regime, \sxy~decreases monotonically for
$T \geq 4.2$ K, as shown in Figs.\ref{Figure3}(c) and (d). The
temperature dependence of \sxx~and \sxy~clearly demarcates two
regimes: at low temperature ($T \lesssim 4.2$ K), we find a
universal behavior for the entire sample set arising from
localization corrections, while at higher temperatures ($T \gtrsim
4.2$ K), the behavior is completely different and non-universal.
Based on the above result, we argue that analyses of the AHE scaling
relationship in \gma~are only meaningful if they are carried out in
the former low temperature regime. For instance, we obtain
identical and consistent results if we carry out the scaling
analysis using either the extrapolated zero temperature
resistivities (obtained from the $T^{1/3}$ dependence of \rxx~and
\rxy) or using the resistivity value measured at the base
temperature of our dilution refrigerator ($< 100$ mK) which is a
stable, reproducible lowest temperature achievable. In both cases,
the following results are found to be the same. However, scaling
carried out at higher temperatures produces significant departures.

\section{Scaling Analysis of the Anomalous Hall Effect at Low Temperatures}
\begin{figure}
\includegraphics[width=8cm]{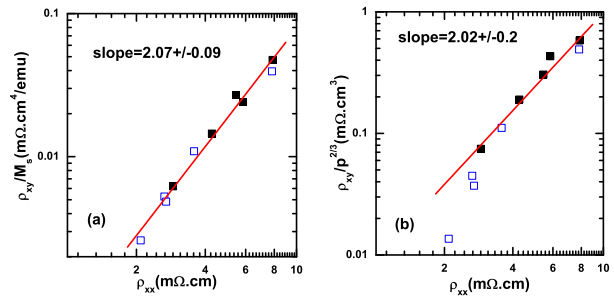}
\caption{\label{Figure4} (a) Zero temperature AH resistivity
normalized by magnetization vs. the zero temperature longitudinal
resistivity; (b) Zero temperature AH resistivity
normalized by magnetization vs. $p^{2/3}$. Linear fits to these log-log plots indicate power law dependence with exponents given by the slopes. Closed (open) symbols represent as-grown (annealed) samples.}
\end{figure}

We now discuss the scaling of our AHE data at low temperature. We
begin with the standard scaling relation of the anomalous Hall
resistivity normalized by the saturated magnetization $M_s$
(anomalous Hall coefficient) $R_s=\rho_{\rm{xy}}/M_s$(Fig.
\ref{Figure4}(a)). A log-log plot of $R_s$ vs.  \rxx~ shows a linear
dependence indicative of a power law behavior of the form $R_s
\sim\rho_{\rm{xx}}^\alpha$. Linear fits to the data yield an
exponent of $\alpha=2.07\pm0.09$. This implies that within
experimental error $R_s/(\rho_{\rm{xx}}^o)^2=\sigma_{\rm{xy}}^o/M_s$
is independent of the variation in the hole carrier density $p$ and
disorder in the sample set under consideration. However, unlike
metallic systems where the above result can perhaps be directly
interpreted as a manifestation of the side jump mechanism, the
situation in \gma~is more complex and requires further examination.
This is because the variation in Mn composition between different
samples and the annealing of a given sample results in a
simultaneous variation of the magnetization, the hole carrier
density and the point defect density.

In ferromagnetic metals, the magnetization arises directly from the
spin imbalance in the conduction band, while the AH voltage arises
from spin-dependent scattering and is directly proportional to the
magnetization. In contrast, \gma~is not a band ferromagnet and the
magnetization is due to the ferromagnetically coupled localized
moments on the substitutional Mn atoms. However, the valence band in
\gma~has a high degree ($\sim 90\%$) of spin polarization
\cite{Braden:2003rx,Panguluri:2005jy}. Consequently, if we assume
that the net spin imbalance is proportional to the exchange field
experienced by the carriers and hence the magnetization
($p_{\uparrow}-p_{\downarrow}\propto M_s$), we encounter a situation
similar to a metal as far as the AHE is concerned. Thus, the result
shown in Fig. \ref{Figure4}(a) can indeed be interpreted as a
manifestation of the side jump mechanism. We estimate the AH
conductivity for the side jump mechanism using the simplified
expression for a fully spin polarized single band system:
$\sigma_{\rm{xy}}=(e^2/\hbar)(p\Delta_{SJ}/k_F)$\cite{BergerPhysRevB.2.4559}.
Using typical values for the carrier density of \gma~in the range of
$10^{20}-10^{21}/\rm{cm ^3}$, we estimate
$\sigma_{\rm{xy}}^{SJ}\sim13 - 63$ $\Omega^{-1}\rm{cm}^{-1}$ (using
$\Delta_{SJ}=$~0.1nm) which is of same order of magnitude as our
experimental values of $10-22$ $\Omega^{-1}\rm{cm}^{-1}$.

To further explore the picture of the side jump mechanism assuming
full spin polarization of carriers, we examine the scaling behavior
of the AH resistivity. If the side jump mechanism indeed contributes
to the AH conductivity through the expression given earlier, then we
expect that $\rho_{\rm{xy}}^o/p^{2/3} \propto {\rho_{\rm{xx}}^o}^2$ (assuming for simplicity a free electron dependence $k_F=
(3\pi^2 p)^{1/3}$). Figure \ref{Figure4}(b) shows a plot testing this
scaling: excluding the three lowest resistivity samples, a linear
fit to the data yields an exponent of $\alpha \sim 2.02\pm0.2$,
implying both that $\sigma_{\rm{xy}}$~ is independent of disorder
and that the data are consistent with the side jump mechanism. The
fact that the high $p$ samples deviate downwards from the linear
dependence is due to the overestimation of the net spin imbalance in
the single band approximation: strictly speaking, the analysis in
Fig.\ref{Figure4}(b) should be carried out with the net carrier spin
imbalance ($p_{\uparrow}-p_{\downarrow}$) rather than the total
carrier density $p$. From the fitting parameters, we estimate the
side jump length to be $(0.07 \pm 0.04)$~nm. 

Much interest has been generated by the theoretical prediction
\cite{JungwirthPhysRevLett.88.207208} of ``dissipationless'' anomalous
Hall current in \gma, originating from purely band structure effects
even in the absence of impurity scattering. 
Calculations in the disorder-free case successfully predict the qualitative
features of the AHE (the sign and rough magnitude of \sxy) in
different III-V DMS materials. As mentioned earlier, recent
experiments \cite{chun:026601,pu:117208} carried out at high
temperatures ($T\sim10-15$ K) have been interpreted using this
picture of an intrinsic AHE in \gma, albeit with the idealized
assumption that compensation is absent. In our sample set, we estimate
$p$ to be in the range of 10-60\% of the total Mn dopant
concentration, indicating a high degree of compensation due to the
presence of interstitials. The only exception was the annealed
sample of $x=0.078$, with a thickness of 50 nm that is found to be
close to the limit of no compensation. Thus, our data cannot provide a rigorous test of the validity of the intrinsic mechanism scenario.

\section{Conclusions}
To conclude, we have carried out a systematic study of the
temperature dependent longitudinal and anomalous Hall conductivity
in a ferromagnetic semiconductor close to the the metal-insulator
transition. We find that carrier localization rather than
interaction effects provide the dominant quantum correction to \sxx,
while our data show no quantum corrections to \sxy. Further, we find
that the absence of temperature dependence in \sxy~and the scaling
relationship between \rxy~and \rxx~is consistent with the side
jump mechanism for the AHE. However, a more rigorous interpretation of these experimental observations will require more detailed calculations for the
AHE in \gma~that explicitly take into account the high degree of disorder in the
samples. We thank S. Potashnik, K. C. Ku and S. H. Chun
for sample growth. The authors thank the Penn State Center for
Nanoscale Science (funded by NSF under Grant No. DMR-0820404) for
the use of low temperature measurement facilities. This work was
also partially supported by NSF under grant DMR-0801406.


\end{document}